\documentstyle[12pt,html,epsf,psfig]{article}

\begin{document}
\title{MATTERS OF GRAVITY, The newsletter of the APS Topical Group on 
Gravitation}
\begin{center}
{ \Large {\bf MATTERS OF GRAVITY}}\\ 
\bigskip
\hrule
\medskip
{The newsletter of the Topical Group on Gravitation of the American Physical 
Society}\\
\medskip
{\bf Number 25 \hfill Spring 2005}
\end{center}
\begin{flushleft}

\tableofcontents
\vfill
\section*{\noindent  Editor\hfill}

Jorge Pullin\\
\smallskip
Department of Physics and Astronomy\\
Louisiana State University\\
Baton Rouge, LA 70803-4001\\
Phone/Fax: (225)578-0464\\
Internet: 
\htmladdnormallink{\protect {\tt{pullin-at-lsu.edu}}}
{mailto:pullin@lsu.edu}\\
WWW: \htmladdnormallink{\protect {\tt{http://www.phys.lsu.edu/faculty/pullin}}}
{http://www.phys.lsu.edu/faculty/pullin}\\
\hfill ISSN: 1527-3431
\begin{rawhtml}
<P>
<BR><HR><P>
\end{rawhtml}
\end{flushleft}
\pagebreak
\section*{Editorial}

This newsletter is horribly late, you have only the editor to blame.
We celebrate the World Year of Physics with our Topical Group in
stronger shape than ever. We have 729 members, up from 654 last year
and 591 the year before. This makes us the second largest topical
group. We constitute 1.68\% of the APS membership. To put this in
perspective, the smallest division has 1,144 members.

The next newsletter is due September 1st. All issues are available in
the WWW:\\\htmladdnormallink{\protect
{\tt{http://www.phys.lsu.edu/mog}}} {http://www.phys.lsu.edu/mog}\\
The newsletter is  available for
Palm Pilots, Palm PC's and web-enabled cell phones as an
Avantgo channel. Check out 
\htmladdnormallink{\protect {\tt{http://www.avantgo.com}}}
{http://www.avantgo.com} under technology$\rightarrow$science.
A hardcopy of the newsletter is distributed free of charge to the
members of the APS Topical Group on Gravitation upon request (the
default distribution form is via the web) to the secretary of the
Topical Group.  It is considered a lack of etiquette to ask me to mail
you hard copies of the newsletter unless you have exhausted all your
resources to get your copy otherwise.  

If you think a topic should be covered by the newsletter you are
strongly encouraged to contact the relevant correspondent.  If you
have comments/questions/complaints about the newsletter email me. Have
fun.

\bigbreak
   
\hfill Jorge Pullin\vspace{-0.8cm}
\parskip=0pt
\section*{Correspondents of Matters of Gravity}
\begin{itemize}
\setlength{\itemsep}{-5pt}
\setlength{\parsep}{0pt}
\item John Friedman and Kip Thorne: Relativistic Astrophysics,
\item Bei-Lok Hu: Quantum Cosmology and Related Topics
\item Gary Horowitz: Interface with Mathematical High Energy Physics and
String Theory
\item Beverly Berger: News from NSF
\item Richard Matzner: Numerical Relativity
\item Abhay Ashtekar and Ted Newman: Mathematical Relativity
\item Bernie Schutz: News From Europe
\item Lee Smolin: Quantum Gravity
\item Cliff Will: Confrontation of Theory with Experiment
\item Peter Bender: Space Experiments
\item Riley Newman: Laboratory Experiments
\item Warren Johnson: Resonant Mass Gravitational Wave Detectors
\item Stan Whitcomb: LIGO Project
\item Peter Saulson: former editor, correspondent at large.
\end{itemize}
\section*{Topical Group in Gravitation (GGR) Authorities}
Chair: Jim Isenberg; Chair-Elect: Jorge Pullin; Vice-Chair: 
\'{E}anna Flanagan;
Secretary-Treasurer: Patrick Brady; Past Chair: John Friedman;
Delegates:
Bei-Lok Hu, Sean Carroll,
Bernd Bruegmann, Don Marolf, 
Gary Horowitz, Eric Adelberger.
\parskip=10pt

\vfill
\pagebreak

\section*{\centerline {
Message from the Chair}}
\addtocontents{toc}{\protect\medskip}
\addtocontents{toc}{\bf GGR News:}
\addcontentsline{toc}{subsubsection}{\it
Message from the Chair, by Jim Isenberg}
\begin{center}
Jim Isenberg, University of Oregon
\htmladdnormallink{jim-at-newton.uoregon.edu}
{mailto:jim@newton.uoregon.edu}
\end{center}

I hope you're not already sick of hearing about Einstein and the 100th 
anniversary of his ``miracle year". We're only one month into it, and it 
remains our job to tell the public how important and revolutionary his 
work of 1905 (and 1915) has turned out to be for our understanding of 
how the universe works. The up and running Speakers Program is doing a 
great job of helping us to convey this to the public. (We thank all of 
those helping with this program, especially Danika Mogilska and Richard 
Price). The special evening session (Einstein's Legacy: What We Know 
and Don't Know) at the April APS meeting in Tampa will help reinforce 
this message in the minds of other  physicists as well.

Of course one way we can view the present Einstein hoopla is as a sort 
of warm up for the BIG centenary celebration in 2015. It is fun to 
think about what we might know about gravitation by then, 100 years 
after Einstein published his paper introducing General Relativity. We 
all hope that detecting gravitational radiation will be relatively 
routine by  2015. Might we even hope that numerical simulations will 
catch up by then? We should know more and more about the very early 
cosmos, and the seeds of clumping into galactic structure. On the other 
end of cosmology, might we have a good model for the apparent 
acceleration of cosmic expansion? It would be great to know a lot more 
than we do about the interface of gravitation and quantum phenomena. 
Likely it is too optimistic to expect an observable manifestation of 
this interface. On the mathematical side, can we hope to have resolved 
whether cosmic censorship (in either the weak or strong form) is true 
for Einstein's gravitational field equations?

Regardless of your attitude on centenaries, this one gives us a good 
excuse to read some of the original papers which Einstein wrote in 1905 
(and later). Just this past week, I reread ``On the Electrodynamics of 
Moving Bodies", and ``Does the Inertia of a Body Depend upon its Energy 
Content". I wouldn't recommend them to Oprah's  Book Club (then again, 
why not?) but I found them to be very interesting reading.

One other anniversarial type note: This is the 10th anniversary of the 
founding of our Topical Group in Gravitation. I think it is fair to say 
that the group has been very successful in advocating and publicizing 
work in gravitational physics. Those of us who agree should give a very 
big thank you to Beverly Berger, whose inspiration and dogged efforts 
ten years ago are responsible for the existence of GGR.

One mark of our success is the increasing number and geographical range 
of our signature ``regional meetings". Starting with the annual spring 
Pacific Coast Gravity Meeting (first held 20 years ago) these have 
spread to the annual fall Midwest Gravity Meeting (first held in 1991), 
the annual spring Eastern Gravity Meeting (first held in 1997), and now 
the newly inaugurated Gulf Coast Gravity Meeting (to be held this 
month). For those of you haven't attended one of these, i highly 
recommend them as a great way to learn about new results in our field, 
and a great way to introduce graduate students to the world of 
gravitational physics research.

\vfill
\eject
\section*{\centerline {
Einstein@Home: a mega-computer for gravitational waves}}
\addcontentsline{toc}{subsubsection}{\it
Einstein@Home, by Bernard Schutz}
\begin{center}
Bernard Schutz, Albert Einstein Institute
\htmladdnormallink{schutz-at-aei.mpg.de}
{mailto:schutz@aei.mpg.de}
\end{center}

With the help of the American Physical Society, the gravitational wave
community is hoping to enlist home computers all over the world in the
search for gravitational waves. The initiative, named Einstein@Home, is
one of APS's World Year of Physics 2005 projects. After the official
release of the software in the first quarter of 2005, anyone will be
able to visit the APS website
\htmladdnormallink{http://www.physics2005.org/}
{http://www.physics2005.org/} and download a screensaver that will
enable any idle computer to become a part of the global gravitational
wave data analysis network.

Einstein@Home is being developed by a team of scientists and
programmers from the LIGO Scientific Collaboration (LSC), led by Bruce
Allen of the University of Wisconsin Milwaukee (UWM). The idea for the
project arose in discussions in 2003 between James Riordon of the APS
and members of the gravitational physics community. Riordon wanted to
use the ``Einstein Year" to provide some practical help to physics, not
just public relations, and he thought that a screensaver could be the
ideal vehicle.  The GW community certainly needs practical help: the
sensitivity of the search for gravitational wave pulsars will be
limited by the available computer power, and even the several
teraflops of cluster computers available within the LSC can make only
a small dent in the problem.

Einstein@Home could make a real breakthrough in the available computer 
power. Even in the preliminary testing phase of the software, enough 
users signed up for it that it could deliver more CPU cycles than any 
other LSC computer. The screensaver is designed to give the computer's 
owner a sense of participating in an important project. It displays a 
rotating globe of the constellations, on which are shown all the known 
pulsars, the current sidereal locations of the LIGO and GEO600 detectors 
(whose data will be analyzed), and the place on the sky where the 
computer is currently doing its search for pulsars. Each computer gets a 
small amount of data from an Einstein@Home server, does the analysis, 
and returns the result. Even if the computer is temporarily disconnected 
from the internet (say, a laptop PC), the analysis will be completed and 
the software will wait for the next opportunity to update itself. Users 
get feedback about how much they have contributed to the effort, and 
they can even join teams that compete to provide more and more cycles!

It is no coincidence that the project's name resembles that of 
SETI@Home. The SETI project (Search for Extraterrestrial Intelligence) 
originated the screensaver-for-science idea, and it has so far managed 
to engage hundreds of thousands of computers in the analysis of short 
stretches of radio telescope data for possible non-random signals. The 
SETI software inspired an open-source product, called BOINC, written by 
SETI@Home developer David Anderson. Einstein@Home is based on the BOINC 
tools. In fact, the Einstein@Home team has made significant 
contributions to BOINC itself. Scientists from LIGO, UWM, and the Albert 
Einstein Institute in Germany have participated in the project.

The Einstein@Home software will run on PCs, Macs, and Linux machines. 
Bruce Allen is hoping not only that members of the general public will 
catch GW fever and sign on, but also that university groups will install 
the screensaver in their computer instruction labs and on their 
departmental workstations. Even typical computers sold in today's mass 
market deliver performance within factors of 5 or better of the chips 
used in many high-performance clusters, so most university physics 
departments can deliver a good fraction of the dedicated computer power 
at any of the LSC computer installations. If Einstein@Home can achieve 
one hundred thousand users (which certainly seems possible), it might 
well turn out that the first gravitational wave source to be discovered 
will be a pulsar found by someone's home PC!
\section*{\centerline {
We hear that...}}
\addcontentsline{toc}{subsubsection}{\it
We hear that..., by Jorge Pullin}
\begin{center}
Jorge Pullin, LSU
\htmladdnormallink{pullin-at-lsu.edu}
{mailto:pullin@lsu.edu}
\end{center}

The SIGRAV (Italian Society of General Relativity and Gravitation) has
awarded the 2004 Amaldi Medal to {\em Roger Penrose}. The Amaldi medal
is a European Prize for Gravitational Physics.  It is awarded
biannually and recognizes an European scientist who has given
outstanding contributions to general relativity and gravitational
physics.

Dieter Brill was elected vice-chair of the Topical Group.
Vern Sandberg was elected secretary/treasurer. Vicky Kalogera and Steven
Penn were elected to the executive committee.

GGR members Larry Ford, Jacqueline Hewitt, Ted Jacobson, Alan Kostelecky, 
Corinne Manogue, Ho Jung Paik, John Price were elected fellows of the APS.

Hearty Congratulations!

\section*{\centerline {
100 Years ago}}
\addcontentsline{toc}{subsubsection}{\it  
100 Years ago, by Jorge Pullin}
\parskip=3pt
\begin{center}
Jorge Pullin
\htmladdnormallink{pullin-at-lsu.edu}
{mailto:pullin@lsu.edu}
\end{center}
German and English versions of Einstein's 1905 paper
``On the electrodynamics of moving bodies'' are available
at 
\htmladdnormallink{http://www.phys.lsu.edu/mog/100}
{http://www.phys.lsu.edu/mog/100}
\vfill
\eject

\section*{\centerline {
What's new in LIGO}}
\addtocontents{toc}{\protect\medskip}
\addtocontents{toc}{\bf Research Briefs:}
\addcontentsline{toc}{subsubsection}{\it  
What's new in LIGO, by David Shoemaker}
\begin{center}
David Shoemaker, LIGO-MIT
\htmladdnormallink{dhs-at-ligo.mit.edu}
{mailto:dhs@ligo.mit.edu}
\end{center}

Since the last MOG, the LIGO Laboratory and more broadly the LIGO
Scientific Collaboration (LSC) has been working on technical
issues in both data analysis and in instrument science. This note
will concentrate on the instruments, complementing the last MOG
LIGO report which described the first observation publications.

{\bf Initial LIGO}

The LIGO interferometers, installed in the observatories in
Livingston, Louisiana, and Hanford, Washington, have interleaved
observation with commissioning over the past few years. Since
completing the S3 run in January of 2004, all the interferometers
have been going through a mixture of tuning and the addition of
new elements to bring them to the desired sensitivity.

An important step forward has been the commissioning of the
Hydraulic External Pre-Isolator, at the Livingston observatory.
This system was originally designed as an element of the Advanced
LIGO seismic isolation system, but was pressed into early
application to reduce the excessively large ground motion in the
0.2 - 10 Hz band at Livingston. It is an active seismic isolation
system, using inertial sensors and actuators in all six degrees of
freedom to reduce the motion of the structure supporting the
original seismic isolation `stack'. It delivers about a factor of
ten reduction in motion -- enough to permit the Livingston
interferometer to lock during the day and through the passage of
trains on a nearby track. This should increase the uptime of the
detector, and also allows commissioning during the day.

Another new element in the interferometers is a Thermal
Compensation System. This is again an Advanced LIGO element
brought to bear on initial LIGO. The notion is to deliver heat to
the interferometer optics to change their focal length (via
$dn/dt$ ). This can be used on `cold' optics to compensate for an
initially slightly incorrect radius of curvature, and/or to
compensate for excessive focusing in a `hot' optic already
distorted by the main sensing laser beam; Gaussian or `doughnut'
profiles, or more complex forms for spatially varying absorption,
are possible. Using the TCS, significant improvements in the
quality of the modulation sidebands have been made along with
reductions in feedthrough of oscillator phase noise. We are still
installing the TCS system, and still learning how to use it, but
it has already and will clearly lead to further reduced shot noise
and other sensing noises.

A variety of other details have been worked to reduce the noise
contributions. Higher gain control loops for the `auxiliary'
servo loops, improved methods to balance the actuators on the
suspensions, modifications of the filtering in photo-detector
amplifiers, and lower-noise oscillators for the
modulation/demodulation systems have all helped. The best
sensitivity to date is in the Hanford 4k instrument, and is shown
in Figure 1. The sensitivity is within a factor of two of the goal
across the entire target sensitivity of the instrument, and we
understand well the remaining limits. We are working on
propagating the improvements to all interferometers.

\begin{figure}[htbp]
  \psfig{file=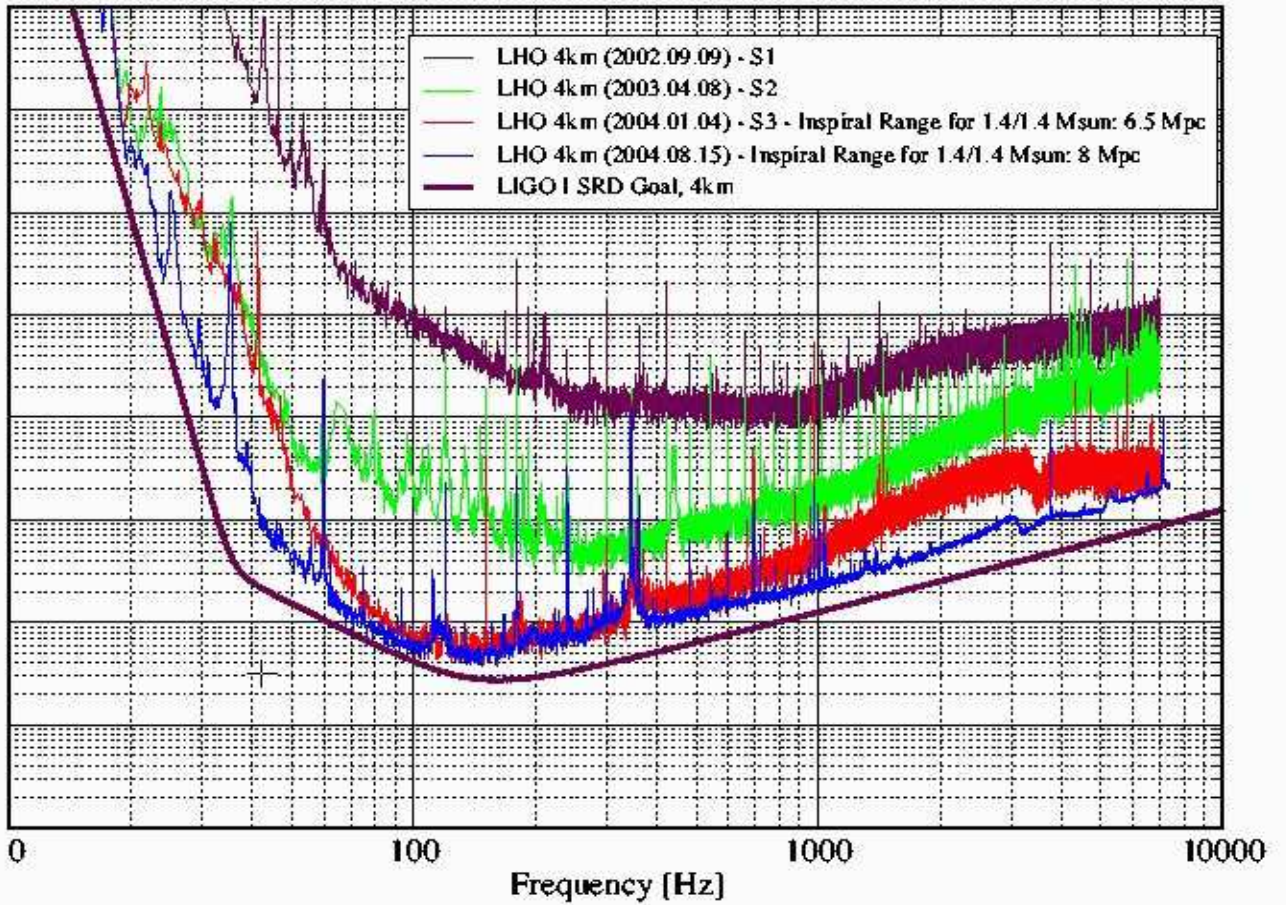}
  \caption{ Caption: The strain sensitivity of the LIGO Hanford
    4km gravitational wave detector, showing its evolution with
    continued commissioning. The bottom-most measured curve dates from
    August, 2004. The smooth line at the bottom is the goal for the
    sensitivity of the instrument, as laid out in the Science
    Requirements Document (`SRD').  (LIGO G040439-00)}
  
\end{figure}

The plan for further data collection and improvements calls for a
one-month observation run, S4, to start in early 2005, followed by
a push to bring all the instruments to the goal for the initial
LIGO instruments. Then the S5 run, currently planned to start late
in 2005, is targeted to collect one year of integrated observation
with the initial LIGO detector.
\vfill\eject
{\bf Advanced LIGO}

The other significant effort in instrument science is to bring
Advanced LIGO forward. An important milestone was passed in
October, when the National Science board reviewed the Advanced
LIGO proposal. They recommended to the director of the NSF that
Advanced LIGO be funded as requested. There are certainly
significant hurdles yet to be passed before funding is received,
but this is a necessary and very important step toward the
realization of Advanced LIGO.

A number of the subsystems have made nice advances recently. The
pre-stabilized laser, led by the Albert Einstein Institut
Hannover, saw their partners Laser Zentrum Hannover achieve the
required 200 watts of laser power from the prototype laser power
head for Advanced LIGO. The high-power test facility at Livingston
came on line, and the Input Optics subsystem led by the University
of Florida started tests of modulator and isolator materials at
realistic power levels. Efforts in the LIGO Laboratory included
optical coating development, which explored parameter space of
dopants and found lower mechanical losses, important for the
thermal noise. Both the requirements for optical losses and
uniformity for the thermal compensation, and successes in making
coatings meeting them, were realized. A full mode-cleaner style
prototype from the Caltech suspension group was installed at the
MIT LASTI facility and characterized, and the quadruple suspension
development, led by the UK/Glasgow with lots of Caltech/MIT
participation, moved forward. The prototype isolation system at
Stanford, with LSU's leadership, was prepared and then operated in
vacuum with new control laws. The 40m interferometer configuration
testbed at Caltech was able to lock all degrees of freedom,
leading the way to tests of the locking of Advanced LIGO and
comparison with models.

One significant point to mention is the choice of substrate for
Advanced LIGO's test mass optics. We had been looking very
carefully since 2001 at two materials: fused silica, which is the
traditional material for fine optics, and used in initial LIGO;
and sapphire, a very hard, high density, low-mechanical loss
material. Careful consideration of both the performance measures
(e.g., they exhibit different thermal noise `signatures', and
different net noise levels for a given coating thermal noise), and
practical questions (e.g., the ability to manufacture and install
complete systems on a schedule) were taken into account.
Curiously, the cost for either option was the same -- so this was
not a net criterion. The Lab has adopted the recommendation from
this study group to use fused silica as the baseline material, and
this allows the quad suspension group to move forward with
dimensions and density for the test mass. A  closing note on this
choice is that, since the internal thermal noise is considered to
be quite low for fused silica, any extra improvements in coating
thermal noise will lead to similar improvements in the Advanced
LIGO sensitivity. A nice challenge!

The coming year will see further full-scale prototyping and
testing, development of readout systems for the interferometer
testbed, development of the complete Advanced LIGO model in the
`e2e' package, and further progress in the other subsystems. And,
we hope, good news on the Advanced LIGO funding timescale.

\vfill
\eject

\section*{\centerline {
Frame-dragging in the news in 2004}}
\addcontentsline{toc}{subsubsection}{\it  
Frame-dragging in the news in 2004, by Cliff Will}
\begin{center}
  Clifford Will\footnote{Full disclosure: the author chairs NASA's
    Science Advisory Committee for Gravity Probe-B} , Washington
  University, St. Louis 
\htmladdnormallink{cmw-at-howdy.wustl.edu}
  {mailto:cmw@howdy.wustl.edu}
\end{center}

Frame-dragging made headlines twice during 2004.  First, on April
20, came the long-awaited launch of Gravity Probe-B, the joint project of
NASA, Stanford University and Lockheed-Martin to measure the dragging of
inertial frames (Lense-Thirring effect), using an array of orbiting
gyroscopes (see Bill Hamilton's article in MOG, Fall 2004).  
By the middle of August, the mission had completed the
commissioning and calibration phase, and commenced science operations.  Now at
the mid-point of the 10-month science phase, the spacecraft and 
instruments are performing as expected [1].
It is too early to know whether the
relativistic effects are being measured in the amount predicted by general
relativity, because an important calibration of the instrument exploits the
effect of the aberration of starlight on the pointing of the on-board
telescope toward the guide star (IM Pegasi), and completing this
calibration requires 
the full mission data set.
In addition, part of the measured effect includes the motion of
the guide star relative to distant inertial frames.  This  is
being measured separately by Irwin Shapiro's group at Harvard/SAO, 
using very long baseline interferometry (VLBI), and the results of those
VLBI
measurements will be strictly embargoed until the GPB team has completed its
analysis of the gyro data.

Meanwhile, on October 21, Ignazio Ciufolini and Erricos Pavlis made
science headlines with a paper in {\em Nature}, in which they claimed to have
measured frame-dragging to between five and 10 percent [2],
using laser ranging to the Earth-orbiting satellites LAGEOS I and II.  

This is not the first report of a measurement of frame dragging using the
LAGEOS satellites.  In 1998 and 2000, Ciufolini and colleagues reported
measurements of the relativistic effect with accuracies ranging from 20 to
30 percent [3,4,5].  What makes this newest report
different from the rest?

The idea behind the LAGEOS experiment is to measure the precession
of the orbital plane caused by the
dragging of inertial frames.  For the LAGEOS satellites, the precession is
about 31 milliarcseconds (mas) per year.  The satellites, launched mainly
for geophysical purposes, are massive spheres studded with laser
retro-reflectors, and as such are not as strongly 
affected by atmospheric drag and
radiation pressure as are complex satellites with solar panels and antennae,
and can also be tracked extremely accurately using laser ranging.

Unfortunately, Newtonian gravity makes a whopping $126^{\rm o} \, {\rm
yr}^{-1}$ contribution to the 
precession.  This haystack 
must be subtracted off,
in order to find the relativistic needle buried within.  The Newtonian
precession depends primarily on the so-called  even zonal harmonics $J_n$ of the
Earth's gravity field, with $J_2$, $J_4$, $J_6 \, \dots$ contributing 
in ever decreasing
amounts.   These moments have been measured over the years using a variety of
Earth-orbiting satellites, but have never been known accurately enough to
permit a simple subtraction of the Newtonian precession.  

In their earlier work, Ciufolini {\em et al.} tried an alternative
method.  Noting that the orbit of LAGEOS II had a small eccentricity,
they argued that, if one measured the two precessions together with
the perigee advance of LAGEOS II, all of which depend on frame
dragging and the zonal harmonics, and if one adopted the existing
values of the harmonics for $n=6$ and higher, then one could use the
three observables to measure the two poorly known $J_2$ and $J_4$, and
the unknown relativity parameter.  This was the basis of the results
presented in Refs. [3,4,5].  Unfortunately, the perigee precession is
strongly affected by non-gravitational perturbations, and so it is
difficult to assess the errors reliably.  A number of experts argued
that the 20 to 30 percent errors assigned by Ciufolini {\it et al.}
were too small by factors as high as five [6,7].

But then along came CHAMP and GRACE.  Europe's CHAMP (Challenging
Minisatellite Payload) and NASA's GRACE (Gravity Recovery and Climate
Experiment) missions, launched in 2000 and 2002, respectively, use
precision tracking of spacecraft to measure variations in Earth's
gravity on scales as small as several hundred kilometers, with
accuracies as much as ten times better than had been obtained
previously.  GRACE consists of a pair of satellites flying in close
formation (200 kilometers apart) in polar orbits.  Each satellite has
on-board accelerometers to measure non-gravitational perturbations,
satellite to satellite K-band radar, to measure variations in the
Earth's gravity gradient on short scales, and GPS tracking to measure
larger scale variations in Earth's gravity.

With the dramatic improvements in $J_n$ obtained 
by CHAMP and GRACE, Ciufolini could now treat $J_4$ and above as known (well
enough), 
drop the troublesome perigee advance, and use the two LAGEOS precessions to
determine $J_2$ and the relativity parameter.  This is what
Ciufolini and Pavlis reported in the recent {\em Nature} paper
[2].

While  all this is valid in
principle, the big question is the treatment of errors.  
Iorio [8] has criticized the error analysis on a number of grounds,
including (i) adopting one GRACE/CHAMP Earth solution for the analysis,
rather than analyzing many solutions for the zonal harmonics and seeing how the
relativity parameter varies; (ii) inadequate treatment of correlations
among the zonal harmonics in the GRACE/CHAMP solutions; and (iii)
inadequate treatment of temporal variations in the low-order harmonics
${\dot J}_4$ and ${\dot J}_6$.   Iorio suggests that the $2-\sigma$
errors should be more like 30 percent [9]

With results from GPB not expected until well after the end of the
mission in July, and with this
lingering discussion of errors in the LAGEOS solutions, we may not
have a solid answer about these measurements of frame dragging
before the end of the Einstein year.

{\bf References:}

[1]
The website for Gravity Probe B is at www.einstein.stanford.edu, and gives
regular updates on the performance of the instruments and spacecraft, as
well as information about how the experiment is designed.

[2]
I. Ciufolini and E. C. Pavlis,  
A confirmation of the general relativistic prediction
of the Lense-Thirring effect,
{\em Nature} {\bf 431}, 958 (2004).

[3]
I. Ciufolini, F. Chieppa, D. Lucchesi and F. Vespe,
Test of Lense - Thirring orbital shift due to spin,
{\em Class. Quantum Gravit.} {\bf 14}, 2701 (1997).

[4]
I. Ciufolini, E. C. Pavlis, F. Chieppa, E. Fernandex-Vieira and P.
P\'erez-Mercader, 
Test of General Relativity and Measurement of the Lense-Thirring
Effect with Two Earth Satellites,
{\em Science} {\bf 279}, 2100 (1998).

[5]
I. Ciufolini,
The 1995-99 measurements of the Lense-Thirring effect using laser-ranged
satellites, {\em Class. Quantum Gravit.} {\bf 17}, 2369 (2000).

[6]
J. C. Ries, R. J. Eanes, B. D. Tapley and G. E. Peterson,
Prospects for an improved Lense-Thirring test with SLR and the GRACE
gravity mission, 
in {\em Proceedings of the 13th International Workshop on Laser Ranging:
Science Session and Full Proceedings CD-ROM},
edited by R. Noomen, S. Klosko, C. Noll, and M. Pearlman,
NASA/CP-2003-212248 (2003);
available online at\\
\htmladdnormallink
{http://cddisa.gsfc.nasa.gov/lw13/lw\_proceedings.html\#science}
{http://cddisa.gsfc.nasa.gov/lw13/lw\_proceedings.html\#science}.

[7]
L. Iorio,
Some comments on the recent results about the measurement of the
Lense-Thirring effect in the gravitational field of the Earth with
the LAGEOS and LAGEOS II satellites, preprint
\htmladdnormallink{gr-qc/0411084}
{http://arxiv.org/abs/gr-qc/0411084}.

[8]
L. Iorio,
Some comments about a recent paper on the measurement of the general
relativistic Lense-Thirring effect in the gravitational field of the
Earth with the laser-ranged LAGEOS and LAGEOS II satellites, preprint
\htmladdnormallink{gr-qc/0410110}
{http://arxiv.org/abs/gr-qc/0410110}.

[9]
Similar comments were made by Ries {\em et al.} [6] in reference to the
1998 analysis of [4]

\vfill
\eject

\section*{\centerline {
Cosmic (super)strings and LIGO}}
\addcontentsline{toc}{subsubsection}{\it  
Cosmic (super)strings and LIGO, by Xavier Siemens}
\begin{center}
Xavier Siemens, University of Wisconsin-Milwaukee
\htmladdnormallink{siemens-at-gravity.phys.uwm.edu}
  {mailto:siemens@gravity.phys.uwm.edu}
\end{center}

Through much of the last two and a half decades, cosmic strings were
of great interest to the cosmology and high energy physics
communities. Unlike other simple topological defects, such as
monopoles and domain walls, strings do not cause cosmological
disasters.  Indeed, cosmic strings formed at the GUT scale would lead
to cosmological density perturbations of the right amplitude to seed
the formation of galaxies and clusters. Thus, cosmic strings became a
leading candidate for structure formation. For a review see
[1].

Cosmic strings were also appealing because their cosmological
evolution (at least the gross features) turned out to be quite simple.
Regardless of the details of the initial conditions, a string network
in an expanding universe will quickly evolve toward an attractor
solution called the ``scaling regime''. In this regime, the energy
density of the string network becomes a small constant fraction of the
radiation or matter density, and the statistical properties of the
system, such as the correlation lengths of long strings and average
sizes of loops, scale with the cosmic time.

The attractor solution is possible due to reconnections, which for
field theoretic strings, essentially always occur when two string
segments meet. Reconnections produce cosmic string loops, which in
turn decay by radiating gravitationally. This takes energy out of the
string network, converting it to gravitational waves. If the density
of string in the network becomes large, then strings will meet more
often, producing extra loops. The loops then decay gravitationally,
removing the surplus energy from the network.  If, on the other hand,
the density of strings becomes too low, strings will not meet to often
enough to produce loops, and their density will start to grow. Thus,
the network is driven toward an equilibrium.

During the 1990s, cosmic microwave background data showed that strings
could not give rise to the density fluctuations that seed structure
formation. These observations placed upper limits on the string
tension below the GUT scale, relegating strings to (at most) a
sub-dominant role in the seeding of structure formation. As a result
of these discoveries, the cosmology community's interest in cosmic
strings dwindled through the late 1990s, and into the
millennium. Recently, however, a few developments have contributed to
a resurgence of interest in cosmic strings.

In 2000, Damour and Vilenkin found that cosmic strings
could lead to the production of sizable gravitational wave
bursts [2]. These bursts may detectable with first generation
ground-based interferometric gravitational wave detectors, such as
LIGO and VIRGO, at design sensitivity. Remarkably, they found values
of the string tension that would result in a measurable signal, that
are below the upper limits placed by cosmic microwave background
observations.

The bursts we are most likely to be able to detect are produced at
cosmic string cusps. These are regions of string which acquire
phenomenal Lorentz boosts, and emit a powerful burst of gravitational
waves in the direction of motion of the string. The formation of cusps
on cosmic string loops and long strings is generic, and their
gravitational waveforms simple and robust [3].

More recently, Jones, Stoica and Tye [4], and Sarangi and Tye
[5] realized that string theory inspired inflation scenarios
lead to the production of cosmic strings. Thus, the very exciting
possibility arises [6] that a certain class of string
theories may have consequences observable in the near future: Just
like ordinary field theoretic strings, the cosmic superstrings formed
could lead to the production of a detectable gravitational wave
signal.

Fortunately, much of what was learned about the evolution of field
theoretic cosmic string networks can be applied to the evolution of
cosmic superstrings. Aside from the possibility of forming more than
one type of string, the most significant difference is that cosmic
superstring interactions are probabilistic. Pairs of strings do not
always reconnect when they meet. Furthermore, strings in higher
dimensional spaces may more readily avoid intersections
[7]. The net effect is to lower the reconnection
probability.  If there is only one type of string, the network still
enters a scaling regime [8], albeit at a density higher by a
factor inversely proportional to the reconnection probability
[9]. It turns out that the smaller reconnection probability
of superstrings actually {\it increases} the chances of detection
through the production of gravitational wave bursts [9]

Finally, there are direct observations that suggest a gravitational
lens produced by a long cosmic string [10], as well as an
oscillating cosmic string loop [11].

The LIGO Scientific Collaboration is currently involved in the
development of a templated search for bursts from strings. At the end
of February 2005, the collaboration plans to start its fourth science
run (S4). The interferometers are within factors of a few from design
sensitivity. Although with current sensitivities a detection seems
unlikely, it may become possible to place constraints on the types of
fundamental particle theories that describe our world.

{\bf References:}

[1] A. Vilenkin and E.P.S Shellard, 
Cosmic strings and other Topological Defects.
Cambridge University Press, 2000; M. Hindmarsh and T.W.B. Kibble, 
Rept. Prog. Phys. 58 (1995) 477.

[2]  T. Damour, A. Vilenkin, Phys.Rev.Lett. {\bf 85} (2000) 3761; 
T. Damour, A. Vilenkin, Phys.Rev.{\bf D}64 (2001) 064008.

[3] X. Siemens, K.D. Olum, Phys.Rev. {\bf D}68 (2003) 085017.

[4] N. Jones, H. Stoica, S.H.Henry Tye, JHEP 0207 (2002) 051, 
\htmladdnormallink{hep-th/0203163}
{http://arxiv.org/abs/hep-th/0203163}.

[5] S. Sarangi, S.H.Henry Tye, Phys.Lett.{\bf B} 536 (2002) 185, 
\htmladdnormallink{hep-th/0204074}
{http://arxiv.org/abs/hep-th/0204074}.

[6] J. Polchinski, 
\htmladdnormallink{hep-th/0410082}
{http://arxiv.org/abs/hep-th/0410082}; 
J. Polchinski, 
\htmladdnormallink{hep-th/0412244}
{http://arxiv.org/abs/hep-th/0412244}.

[7] G. Dvali, A. Vilenkin, JCAP 0403 (2004) 010.

[8] N. Jones, H. Stoica, S.H.Henry Tye, Phys.Lett. {\bf B}563 (2003) 6.

[9] T. Damour, A. Vilenkin, 
\htmladdnormallink{hep-th/0410222}
{http://arxiv.org/abs/hep-th/0410222}.

[10] M. Sazhin et al., Mon. Not. Roy. Astron. Soc. {\bf 343} (2003) 353;
M. Sazhin et al., 
\htmladdnormallink{astro-ph/0406516}
{http://arxiv.org/abs/astro-ph/0406516}.

[11] R. Schild et al., 
\htmladdnormallink{astro-ph/0406434}
{http://arxiv.org/abs/astro-ph/0406434}.

\vfill
\eject

\section*{\centerline {
The first gulf coast gravity conference (GC)$^2$}}
\addtocontents{toc}{\protect\medskip}
\addtocontents{toc}{\bf Conference reports:}
\addcontentsline{toc}{subsubsection}{\it  
The first gulf coast gravity conference, by Richard Price}
\parskip=3pt
\begin{center}
Richard Price, University of Texas at Brownsville
\htmladdnormallink{rprice-at-phys.utb.edu}
{mailto:rprice@phys.utb.edu}
\end{center}

There is a pronounced drift of relativity into the region from Texas
to Florida, a range of states that does not fit into the reach of the
other regional meetings. For that reason there is now yet another
regional meeting: The Gulf Coast Gravity Conference.

The first such meeting took place Feb. 11 and 12 at the University of
Texas at Brownsville, organized by Carlos Lousto, and was attended by
relativists from Texas (UT Austin as well as Brownsville), Louisiana
(LSU and LIGO), and Florida (Florida Atlantic University).  As in
other regional meetings, this was a meeting in which everyone got the
same short time to talk, and in which students were giving their first
talk in front of ``outsiders."


The first day's focus was numerical relativity, and showed that the
present state of the field has both convergence of some results and
controversy about others. There was agreement about plunge radiation.
Results reviewed by UTB's Manuela Campanelli from the Lazarus project
(numerical relativity plus perturbation theory) showed about 2.5\% of
the mass-energy and about 12\% of the angular momentum radiated in the
late stage of infall.  Recent results from LSU, reported by Peter
Diener, were in good agreement with the conclusion about radiated
energy and reasonable (all things considered) agreement about angular
momentum.

There was a useful lack of agreement about initial data, however.
Pedro Maronetti of FAU, reporting recent results, showed an efficient
way of doing short term evolutionary tests of initial data to see
whether the thin-sandwich initial data advertised for circular orbits
really does give circular orbits. His recent results, for neutron
stars, were in good agreement with circular orbits.  Results reported
by Peter Diener, from Bowen-York initial data for circular orbits sort
of suggested the opposite for black holes. Is it thin-sandwich vs
Bowen-York, or something else.  It will be interesting to watch how
answers to this question develop.

The UTB numerical relativity team, Joseph Zochlower (fourth order
codes), Mark Hannam (initial data), and Bernard Kelly (reference
frames for Lazarus), gave updates on the many refinements that now
characterize the state of the art in numerical relativity. Steve Lau,
of UTB, reported on work on radiation boundary conditions indicating
that reflections off an outer boundary could be 
eliminated in principle, and probably greatly reduced in practice.

The second day of the conference contained several talks related to my
own recent obsession: an ``intermediate" approximation for black hole
inspiral: an approximation appropriate to the later-than-PN,
too-soon-for-numerical relativity approximation. I showed the
efficiency of a new numerical technique we were using. Chris Beetle
has reported on a general mathematical result for how Kepler's law
would come out of the intermediate approximation, and Mike McLaughlin
showed how some modern applied mathematics could be used to reduce the
computational infrastructure that our method might require.

In the afternoon, Rayesh Nayak described the orbital dynamics that are
the basis of LISA, and we had talks on data analysis by Alexander
Dietz of LSU, and by students Andres Rodriguez (LSU), and Arturo
Jimenez, and Charlie Torres (UTB). The widely varying approaches
described were a reminder of how difficult this problem will be, and
on how open the questions remain.

The prize for the best of an exceptionally good set of student talks
was won by Napoleon Hernandez of UTB for his presentation ``Towards the
computation with Dirac Delta sources for the Teukolsky equation."

\vfill
\eject

\section*{\centerline {
Imagining the future: gravitational wave astronomy}}
\addcontentsline{toc}{subsubsection}{\it  
Imagining the future, by Shane Larson}
\parskip=3pt
\begin{center}
Shane Larson, PennState
\htmladdnormallink{shane-at-gravity.psu.edu}
{mailto:shane@gravity.psu.edu}
\end{center}

On October 27-30, 2004, a group of 64 gravitational wave astronomers
and astronomers from traditional fields of astronomy and astrophysics,
representing 20 different institutions, convened at Penn State for a
workshop to speculate on the future of gravitational wave astronomy.
The purpose of the workshop was to begin a conversation in the
community about the future evolution of the field of gravitational
wave physics and astronomy, and to consider how this new observational
science will fit into the toolbox of modern astronomers.  The active
pursuit of imagining the future, working outside of the confines of
today's vexing research problems, helps to illuminate the road ahead
so the needs of the field can be addressed in advance of the time when
progress will be hindered or facilitated by the nature and quality of
the infrastructure supporting the community as a whole.  This type of
endeavor is typically known as \emph{strategic planning}.

The attendees were asked to consider a time two decades in the
future, when gravitational wave astronomy is an established (but
perhaps still adolescent) observational science that regularly
contributes to our view of the Universe as an active component of
multi-messenger astronomy.  To facilitate discussion and debate
oriented toward considering the future of the field, six questions
were posed:
\begin{itemize}
    \item What will it mean to be a ``gravitational wave astronomer''?
    
    \item What will be the interplay between gravitational wave 
    astronomy and other, now conventional, forms of astronomy?
    
    \item What will be the interplay between instrumentation, 
    observation, and science in the field?
    
    \item What will be the role of individual observatories vs. global 
    networks?
    
    \item What will be the critical technologies used in 
    gravitational wave detection?
    
    \item What infrastructure will best contribute to broad 
    participation, community growth, and the best possible science?
\end{itemize}
After several days of open discussion and debate centered on these 
questions, several key findings emerged.

First, it is likely that in 20 years there may not be people who
self-identify as ``gravitational wave astronomers.''  Instead,
practitioners will become either instrumentalists or astronomers who
tap gravitational wave astronomy as one element in a suite of tools
used to probe the Universe.  Today, the growth of multi-messenger
astronomy is evident from cross-communication between electromagnetic
bands and astroparticle astronomy.  In an era where gravitational wave
observations and detections are routine, it is not unreasonable to
expect that gravitational astronomy will simply be another element
which strengthens our ability to probe high energy astrophysical
systems as part of multi-messenger observing campaigns.

Data products and data access were discussed at great length.
Currently gravitational wave astronomy is operating with data products
in which instrumental noise and astrophysical signal strength are
comparable at best, and data analysis efforts require the expertise of
people intimately familiar with the detectors.  In the future, as
instrumentation evolves and sensitivity increases, this model of data
control will become less desirable.  Data products should become
broadly available to the astronomical community as a constituent of
the multi-spectrum information accessible for research efforts.

International collaborations will play an increasingly important role
in the construction and operation of gravitational wave observatories.
Large collaborations which span the globe already play a vital role in
the community, as readily evidenced by the existing networks of bar
detectors and the emerging network of interferometric observatories.
These networks facilitate the growth of technology, the sharing of
expertise, and increase the capacity to pursue fundamental and large
scale science initiatives.  International networks will only continue
to grow in the future of gravitational wave astronomy.

Over the next two decades, it is desirable to see undergraduate and
graduate astronomy curricula evolve to include observational
gravitational wave science as part of the suite of tools available to
the modern astronomer.  This is not a suggestion to inundate
astronomy curricula with deep courses in general relativity, but to
establish a curriculum which encompasses the fundamental science which
can be learned from gravitational wave observations.  This finding has
been summarized by the mantra ``Less $G_{\mu\nu}$, more Rybicki \&
Lightman!''

Lastly, attention needs to be paid to Research and Development and
budgets established to support emerging technologies.  An agreed upon
rule of thumb for R\&D budgets in growing astronomical fields is
$5-10\%$ of the total budgetary support in the field; estimates at the
workshop suggested gravitational wave astronomy currently expends on
order of $1-2\%$.  Critical technologies which should bear scrutiny as
we advance toward regular gravitational wave observations of the
Universe are high power and high stability lasers, quantum
non-demolition techniques, development of advanced materials, and
computation and data analysis infrastructure.

A white paper summarizing the key findings and open debates left by
the conference is in preparation, and will be posted to
\texttt{arxiv.org} when it is completed.  The program of the
conference and the talks which were presented to encourage discussion
have been posted online at the Center for Gravitational Wave Physics
at Penn State, linked at
\htmladdnormallink{http://cgwp.gravity.psu.edu/events/GWA/}
{http://cgwp.gravity.psu.edu/events/GWA/}.

It was intended that this workshop would be only the first of many
such discussions the community will have with itself over the next
decade.  We strongly encourage everyone to consider these initial
findings and whether they agree or disagree with them, to carry on the
discussions and debates with your colleagues and groups at your home
institutions, and to participate in future strategic planning events
like this.

\vfill
\eject

\section*{\centerline {
    VI Mexican School on Gravitation and Mathematical Physics}}
\addcontentsline{toc}{subsubsection}{\it VI Mexican School, by
  Alejandro Corichi}
\begin{center}
Alejandro Corichi, National Autonomous University of Mexico (UNAM)
\htmladdnormallink{corichi-at-nuclecu.unam.mx}
{mailto:corichi@nuclecu.unam.mx}
\end{center}

The Division of Gravitation and Mathematical Physics of the Mexican
Physical Society organized its VI school, November 21-28th, near Playa
del Carmen, in the Mexican Riviera. These schools take place every
other year and are focused on a special topic. This year the school
was named ``Approaches to Quantum Gravity''. The setting was specially
welcoming, being an all inclusive resort that allowed the participants
to fully enjoy the amenities of the hotel (and the beach), while
focusing on the scientific program of the school.

The idea of the organizers was to have top class representatives from
at least three approaches to quantum gravity deliver lectures for
graduate students and non-experts. The overall opinion is that the
goal was achieved and the school was rather successful. There were
about 100 participant in total, with about half of them students from
both Mexico and abroad.

The speakers were chosen to cover the two main approaches to quantum
gravity, namely string theory and loop quantum gravity, as well as
other approaches and related subjects. On the stringy side the courses
were delivered by R. Kallosh and A. Peet, together with plenary talks
by A. Guijosa and E. Caceres. On the loop side, the invited speakers
for the courses were A. Ashtekar and C.  Rovelli. Ashtekar had to
cancel at the last minute so he was replaced by the author, M.
Bojowald and L. Smolin. There were also plenary talks by M. Bojowald,
A. Perez and L. Smolin, on loop related issues. On the `other
approaches category' there was a course on time-space
non-commutativity by A.P. Balachandran and a course on selected
classical topics by P. Chrusciel. The list of distinguished plenary
speakers also included J. Barrow, A. Linde and R. Wald, who spoke on
variable fundamental constants, inflationary cosmology and QFT on
curved space, respectively.

String Theory was very well represented by R. Kallosh and A. Peet who
delivered set of lectures on ``De Sitter vacua and the String
Landscape'' (Kallosh) and ``Black Holes in String Theory" (Peet).  They
gave a very though overview of recent research results and the cutting
edge research on the subject. R. Kallosh first motivated the need for
a theoretical explanation for the value of the positive cosmological
constant observed. She then described a recent model she has been
working on that involves choosing certain vacua from the string
landscape, that is, from the very large set of possible vacua for the
theory. Kallosh also described recent work on black holes and the
search for the fundamental degrees of freedom of the theory. The set
of lectures delivered by A. Peet were particularly illuminating. She
described the basics of string theory, with some clarifications that
answer some common misconceptions outsiders sometimes have. She
described in detail the D-brane approach to BH entropy, AdS-CFT
correspondence and the newly discovered fuzzball solutions. She ended
the set of lectures with an inspiring challenge for the participants
of the school.

On the loop side of the school, the author introduced the basics
of loop quantum geometry, Bojowald applied the formalism to
cosmological solutions, arriving at the so called loop quantum
cosmology, and Smolin described the challenges and possible
resolutions for the dynamics of the theory. C. Rovelli gave a very
nice set  of lectures on the white-board, motivating and building
loop quantum gravity from scratch. The courses were complemented
by lectures on black holes, spin foams and phenomenological
aspects of loop quantum cosmology.

The day before the school ended, R. Wald moderated a discussion
panel where the theme was Strings {\it vs.} Loops. The discussion
was polite but lively, and questions like ``what are fundamental
degrees of freedom in string theory?" and ``how does one make
sense of spatially smeared operators in LQG?" were discussed. The
general agreement was that the discussion was very civil, but
there was not enough time to discuss some more controversial
issues.

The school ended by an inspiring lecture by Lee Smolin who
discussed various possible experimental indications for the need
of a quantum theory of gravity and a new length scale as a
fundamental entity. For more information on the program of the
School see 

\htmladdnormallink{http://www.nuclecu.unam.mx/\~{}gravit/EscuelaVI}
{http://www.nuclecu.unam.mx/\~gravit/EscuelaVI}.

\end{document}